\documentstyle[aps,prd]{revtex}
\input epsf.sty

\begin{document}
\hbadness=10000 \pagenumbering{arabic}

\preprint{{\vbox{\hbox{ HEP-PH/0202XXX}}}} \vspace{1.5cm}

\title{\Large \bf
Perturbative QCD analysis of $B \rightarrow \phi K^*$ decays}

\date{\today}
\author{\large \bf Chuan-Hung Chen$^{a}$\footnote{Email:
chchen@phys.nthu.edu.tw}, Yong-Yeon Keum$^{b}$\footnote{Email:
yykeum@eken.phys.nagoya-u.ac.jp} and Hsiang-nan
Li$^{a,c}$\footnote{Email: hnli@phys.sinica.edu.tw}}

\vskip1.0cm

\address{ $^{a}$ Institute of Physics, Academia Sinica,
Taipei, Taiwan 115, Republic of China}
\address{ $^{b}$ Department of Physics, Graduate School of Science,
Nagoya University, Nagoya 464-8602, Japan}
\address{ $^{c}$ Department of Physics, National Cheng-Kung University,
Tainan, Taiwan 701, Republic of China}
\maketitle

\begin{abstract}

We study the first observed charmless $B\to VV$ modes, the $B\to\phi K^*$
decays, in perturbative QCD formalism. The obtained
branching ratios $B(B\to\phi K^*)\sim 15 \times 10^{-6}$ are larger than
$\sim 9\times 10^{-6}$ from QCD factorization. The comparison of the
predicted magnitudes and phases of the different helicity amplitudes, and
branching ratios with experimental data can test the power counting
rules, the evaluation of annihilation contributions, and the mechanism of
dynamical penguin enhancement in perturbative QCD, respectively.

\end{abstract}

\vskip 1.0cm

The branching ratios of the penguin-dominated $B\to K\pi$ decays,
about 3-4 times larger than those of the tree-dominated $B\to\pi\pi$
decays, indicate that penguin contributions must be enhanced. This
enhancement can be achieved either by large Wilson coefficients
$C_{4,6}$ associated with the penguin operators in perturbative QCD (PQCD)
\cite{Keum:2001ph,KL,LUY}, or by a large chiral symmetry breaking
scale $m_0$ associated with the kaon in QCD factorization (QCDF)
\cite{BBNS,BBNS2}. The latter mechanism, called chiral enhancement,
corresponds to a characteristic scale of $O(m_b)$, at which we have
$m_0(m_b)\sim 3$ GeV and the smaller Wilson coefficients $C_{4,6}(m_b)$.
The former mechanism, called dynamical enhancement, corresponds to a
characteristic scale of $O(\sqrt{\bar\Lambda m_b})$,
$\bar\Lambda=M_B-m_b$ being the $B$ meson and $b$ quark mass difference,
at which we have $m_0(\sqrt{\bar\Lambda m_b})\sim 1.5$ GeV and the larger
Wilson coefficients $C_{4,6}(\sqrt{\bar\Lambda m_b})\sim 1.5 C_{4,6}(m_b)$.
Recently, we have proposed the $B\to\phi K$ decays as the appropriate
modes to clarify the above issue \cite{L6,CKL}. These modes are not
chirally enhanced, because $\phi$ is a vector meson, and insensitive to
the variation of the unitarity angle $\phi_3$, because they are pure
penguin processes. If the data of the branching ratios $B(B\to \phi K)$
are settled down at values around $10\times 10^{-6}$ \cite{CKL,Mi},
instead of $4\times 10^{-6}$ \cite{CY1,HMW}, the dynamical enhancement of
penguin contributions to charmless nonleptonic $B$ meson decays will
gain a strong support.

Here we argue why the characteristic scale involved in two-body $B$
meson decays must be of $O(\sqrt{\bar\Lambda M_B})$ in PQCD from two
points of view. Consider a two-body nonleptonic decay, in which the two
final-state light mesons move back-to-back with large momenta. The
lowest-order diagram for its amplitude contains a hard gluon attaching
the spectator quark. Intuitively, the spectator quark in the $B$ meson,
forming a soft cloud around the heavy $b$ quark, carries momentum of order
$\bar\Lambda$. The spectator quark on the light-meson side carries momentum
of $O(M_B)$ in order to form the fast-moving light meson with the $u$ quark
produced in the $b$ quark decay. Note that the end-point singularities
from the small spectator momentum on the light-meson side do not exist
in a self-consistent PQCD formalism, because of Sudakov
suppression from $k_T$ and threshold resummations \cite{TLS,WY}.
Based on the above argument, the hard gluon is off-shell by order of
$\bar\Lambda M_B$. This scale characterizes the corresponding
quark-level hard amplitude, which involves the four-fermion decay
vertex. Theoretically, the hard scale $\bar\Lambda M_B$ is essential for
constructing a gauge invariant $B$ meson wave function. This wave function,
though being a nonlocal matrix element, is gauge invariant in the
presence of the path-ordered Wilson line integral. A careful
investigation \cite{L4,NL} shows that the $O(\alpha_s^2)$ diagram
with the second gluon attaching the hard gluon contributes to this
line integral. That is, this diagram contains the soft divergence,
which is factorized into the $B$ meson wave function. This is possible,
only when the hard gluon is off-shell by the intermediate scale
$\bar\Lambda M_B$, rather than by $M_B^2$.

In this work we shall perform a PQCD analysis of the first
observed charmless $B\to VV$ modes, the $B\to\phi K^*$ decays,
which are, similar to $B\to\phi K$, also appropriate for
distinguishing the different penguin enhancing mechanism. Besides,
the $B\to VV$ modes reveal dynamics of exclusive $B$ meson decays
more than the $B\to PP$ and $VP$ modes through the measurement of
the magnitudes and the phases of various helicity amplitudes.
According to the power counting rules defined in \cite{CKL}, the
longitudinal amplitude is leading, and the other two amplitudes
are down by a power of $M_\phi/M_B$ or of $M_{K^*}/M_B$, $M_\phi$
and $M_{K^*}$ being the $\phi$ and $K^*$ meson masses,
respectively. Since the $B\to\phi K^*$ decays are insensitive to
the unitarity angle, the relative phases among the helicity
amplitudes mainly arise from strong interaction. The annihilation
contributions, which can be evaluated unambiguously in our
approach, generate the strong phases. Therefore, comparing the
predicted magnitudes and relative phases among the different
helicity amplitudes, and the predicted branching ratios with
experimental data, we test the power counting rules, the
evaluation of annihilation contributions, and the mechanism of
dynamical penguin enhancement in PQCD, respectively.

The idea of the PQCD factorization theorem for two-body
nonleptonic $B$ meson decays has been reviewed in
\cite{Keum:2001ph,CL,YL}, which is subject to corrections of
$O(\alpha_s^2)$ and $O(\bar\Lambda/M_B)$. In this formalism decay
amplitudes are expressed as the convolutions of the corresponding
hard parts with universal meson distribution amplitudes
\cite{L4,NL}, which are regarded as the nonperturbative inputs.
Because of the Sudakov effects from $k_T$ and threshold
resummations, the end-point singularities do not exist as stated
above. Therefore, PQCD involves inputs less than in QCDF, for
which form factors, meson distribution amplitudes, and infrared
cutoffs for regulating the end-point singularities are all
independent parameters \cite{BBNS,BBNS2}. Strictly speaking, the
infrared cutoffs, signifying important soft contributions to the
nonfactorizable and annihilation amplitudes, imply that the
factorization formulas in QCDF are not self-consistent.

We work in the frame with the $B$ meson at rest, {\it i.e.}, with the
$B$ meson momentum $P_1=(M_B/\sqrt{2})(1,1,{\bf 0}_T)$ in the light-cone
coordinates. Assume that the $\phi$ ($K^*$) meson moves in the plus
(minus) $z$ direction carrying the momentum $P_2$ ($P_3$) and the
polarization vectors $\epsilon_2$ ($\epsilon_3$). The $B\to \phi K^*$
decay rates are written as
\begin{equation}
\Gamma =\frac{G_{F}^{2}P_c}{16\pi M^{2}_{B} }
\sum_{\sigma=L,T}{\cal M}^{(\sigma)\dagger }{\cal M^{(\sigma)}}\;,
\label{dr1}
\end{equation}
where $P_c\equiv |P_{2z}|=|P_{3z}|$ is the momentum of either of the
outgoing vector mesons, and the superscript $\sigma$ denotes the
helicity states of the two vector mesons with $L(T)$ standing for the
longitudinal (transverse) component. The amplitude
$\cal M^{(\sigma)}$ is decomposed into
\begin{eqnarray}
{\cal M}^{(\sigma)}
&=&\epsilon_{2\mu}^{*}(\sigma)\epsilon_{3\nu}^{*}(\sigma)
\left[ a \,\, g^{\mu\nu} + {b \over M_\phi M_{K^*}} P_1^\mu P_1^\nu
+ i{c \over M_\phi M_{K^*}} \epsilon^{\mu\nu\alpha\beta} P_{2\alpha}
P_{3\beta}\right]\;,
\nonumber \\
&\equiv &M_{B}^{2}{\cal M}_{L}+M_{B}^{2}{\cal M}_{N}
\epsilon^{*}_{2}(\sigma=T)\cdot\epsilon^{*}_{3}(\sigma=T) +i{\cal
M}_{T}\epsilon^{\alpha \beta\gamma \rho}
\epsilon^{*}_{2\alpha}(\sigma)\epsilon^{*}_{3\beta}(\sigma)
P_{2\gamma }P_{3\rho }\;,
\end{eqnarray}
with the convention $\epsilon^{0123} = 1$\footnote{This convention
corresponds to $tr(\gamma_5\not a\not b\not c\not d)=
-4i\epsilon^{\alpha\beta\gamma\rho}a_\alpha b_\beta c_\gamma d_\rho$.}
and the definitions,
\begin{eqnarray}
M_B^2 \,\, M_L &=& a \,\, \epsilon_2^{*}(L) \cdot \epsilon_3^{*}(L)
+{b \over M_\phi M_{K^*}} \epsilon_{2}^{*}(L) \cdot P_1 \,\,
\epsilon_{3}^{*}(L) \cdot P_1\;,
\nonumber \\
M_B^2 \,\, M_N &=& a \,\, \epsilon_2^{*}(T) \cdot \epsilon_3^{*}(T)\;,
\label{id-rel} \\
M_T &=& {c \over M_\phi M_{K^*}}\;.
\nonumber
\end{eqnarray}

We define the helicity amplitudes,
\begin{eqnarray}
A_{0}&=&-\xi M^{2}_{B}{\cal M}_{L}, \nonumber\\
A_{\parallel}&=&\xi \sqrt{2}M^{2}_{B}{\cal M}_{N}, \nonumber \\
A_{\perp}&=&\xi M_{\phi} M_{K^*} \sqrt{2(r^{2}-1)} {\cal M }_{T}\;,
\label{ase}
\end{eqnarray}
with the normalization factor
$\xi=\sqrt{G^2_{F}P_c/(16\pi M^2_{B}\Gamma)}$ and the ratio
$r=P_{2}\cdot P_{3}/(M_{\phi}M_{K^*})$. These helicity amplitudes satisfy
the relation,
\begin{eqnarray}
|A_{0}|^2+|A_{\parallel}|^2+|A_{\perp}|^2=1\;,
\end{eqnarray}
following the helicity summation in Eq.~(\ref{dr1}). We also introduce
another equivalent set of helicity amplitudes,
\begin{eqnarray}
H_{0}&=&M^{2}_{B} {\cal M}_L\;,
\nonumber\\
H_{\pm}&=& M^{2}_{B} {\cal M}_{N} \mp  M_{\phi} M_{k^*}
\sqrt{r^2-1}{\cal M}_{T}\;,
\end{eqnarray}
with the helicity summation,
\begin{eqnarray}
\sum_{\sigma}{\cal M}^{(\sigma)\dagger }{\cal M^{(\sigma)}} =
|H_{0}| ^{2}+|H_{+}|^{2} + | H_{-}|^{2}\;.
\end{eqnarray}

The $B\rightarrow \phi K^{*}$ decays involve the emission and annihilation
topologies, both of which are classified into factorizable diagrams,
where hard gluons attach the valence quarks in the same meson, and
nonfactorizable diagrams, where hard gluons attach the valence quarks in
different mesons. The amplitudes are written as
\begin{eqnarray}
{\cal M}_{H}&=&f_{\phi }V^{*}_{t}F_{He}^{( s) } +V^{*}_{t}{\cal
M}_{He}^{(s) } +f_{B}V^{*}_{t}F_{Ha}^{(d) } +V^{*}_{t}{\cal
M}_{Ha}^{(d) }\;,
\\
{\cal M}_{H}&=&f_{\phi }V^{*}_{t}F_{He}^{( s) } +V^{*}_{t}{\cal
M}_{He}^{(s) } +f_{B}V^{*}_{t}F_{Ha}^{(u) } +V^{*}_{t}{\cal
M}_{Ha}^{(u) } -f_{B}V^{*}_{u}F_{Ha} -V^{*}_{u}{\cal M}_{Ha}\;,
\label{dec}
\end{eqnarray}
for the $B_d^0\to \phi K^{*0}$ and $B^+\to \phi K^{*+}$ modes,
respectively, where the subscript $H=L,N,T$ denotes the different
helicity amplitudes, $e$ ($a$) denotes the emission (annihilation)
topology, and $V_{q}=V_{qs}^{*}V_{qb}$ are the products of the
Cabibbo-Kobayashi-Maskawa (CKM) matrix elements. The hard parts
for the factorizable amplitudes $F$ and for the nonfactorizable
amplitudes $\cal M$ are derived by contracting the following
structures to the lowest-order one-gluon-exchange diagrams,
\begin{eqnarray}
& &\frac{1}{\sqrt{2N_c}}(\not P_1+M_B)\gamma_5\Phi(x,b)\;,
\\
& &\frac{1}{\sqrt{2N_c}}\left[M_\phi\not \epsilon_2(L)\Phi_\phi(x)
+\not\epsilon_2(L)\not P_2 \Phi_{\phi}^{t}(x)
+M_\phi I\Phi_\phi^s(x)\right]\;,
\label{lpf}\\
& &\frac{1}{\sqrt{2N_c}}
\left[M_\phi\not \epsilon_2(T)\Phi_\phi^v(x)+
\not\epsilon_2(T)\not P_2\Phi_\phi^T(x)
+\frac{M_\phi}{P_2\cdot n_-}
i\epsilon_{\mu\nu\rho\sigma}\gamma_5\gamma^\mu\epsilon_2^\nu(T)
P_2^\rho n_-^\sigma \Phi_\phi^a(x)\right]\;,
\label{spf}\\
& &\frac{1}{\sqrt{2N_c}}\left[M_{K^*}\not \epsilon_3(L)\Phi_{K^*}(x)
+\not\epsilon_3(L)\not P_3 \Phi_{K^*}^{t}(x)
+M_{K^*} I\Phi_{K^*}^s(x)\right]\;,
\label{klpf}\\
& &\frac{1}{\sqrt{2N_c}}
\left[M_{K^*}\not \epsilon_3(T)\Phi_{K^*}^v(x)+
\not\epsilon_3(T)\not P_3\Phi_{K^*}^T(x)
+\frac{M_{K^*}}{P_3\cdot n_+}
i\epsilon_{\mu\nu\rho\sigma}\gamma_5\gamma^\mu\epsilon_3^\nu(T)
P_3^\rho n_+^\sigma \Phi_{K^*}^a(x)\right]\;.
\label{kspf}
\end{eqnarray}
where $n_+=(1,0,{\bf 0}_T)$ and $n_-=(0,1,{\bf 0}_T)$ are dimensionless
vectors on the light cone. Equations (\ref{lpf}) and (\ref{spf}) are
associated with the longitudinally and transversely polarized $\phi$
mesons, respectively. The structures associated with the $K^*$ meson are
similar as shown above.

To extract the contributions to the helicity amplitude ${\cal M}_L$,
the following parametrization for the longitudinal polarization vectors
is useful:
\begin{eqnarray}
\epsilon_2(L)=\frac{P_2}{M_\phi}-\frac{M_\phi}{P_2\cdot n_-}n_-\;,\;\;\;\;
\epsilon_3(L)=\frac{P_3}{M_{K^*}}-\frac{M_{K^*}}{P_3\cdot n_+}n_+\;,
\end{eqnarray}
which satisfy the normalization
$\epsilon_2^2(L)=\epsilon_3^2(L)=-1$ and the orthogonality
$\epsilon_2(L)\cdot P_2=\epsilon_3(L)\cdot P_3=0$ for the on-shell
conditions $P_2^2=M_\phi^2$ and $P_3^2=M_{K^*}^2$. We first keep
the full dependence on the light meson masses $M_\phi$ and
$M_{K^*}$ in the momenta $P_2$ and $P_3$. After deriving the
factorization formulas, which are well-defined in the limit
$M_{\phi},M_{K^*}\to 0$, we drop the terms proportional to
$r^{2}_{\phi},\ r_{K^*}^2\sim 0.04$, with the ratios
$r_\phi=M_{\phi}/M_B$ and $r_{K^*}=M_{K^*}/M_B$. Under this
approximation, the expressions of the $\phi$ and $K^*$ meson
momenta are then as simple as
\begin{eqnarray}
P_2=\frac{M_B}{\sqrt{2}}(1,0,{\bf 0}_T)\;,\;\;\;\;
P_3=\frac{M_B}{\sqrt{2}}(0,1,{\bf 0}_T)\;.
\label{pal}
\end{eqnarray}
For the extraction of the helicity amplitudes ${\cal M}_N$ and
${\cal M}_T$, Eq.~(\ref{pal}) and the transverse polarization vectors,
\begin{eqnarray}
\epsilon_2(T) =(0,0,{\bf 1}_T)\;,\;\;\;\;
\epsilon_3(T) =(0,0,{\bf 1}_T)\;,
\end{eqnarray}
can be adopted directly. The explicit factorization formulas are collected
in the Appendix.

The power counting rules in PQCD \cite{CKL} tells that the
factorizable amplitude $F_{Le}$ (corresponding to the $B\to K^*$
transition form factor) is leading, and the other factorizable
amplitudes are at least down by a power of $r_{\phi}$ or
$r_{K^*}$. The nonfactorizable amplitudes $\cal M$ are suppressed
by a power of $\bar\Lambda/M_B$. Hence, the formalism presented in
this work is complete at $O(M_{\phi,K^*}/M_B)$, and subject to
corrections of $O(\bar\Lambda/M_B)$. Equation~(\ref{ase}) then
implies that the helicity amplitude $A_0$ is leading in the
heavy-quark limit, and $A_\parallel$ and $A_\perp$ are
next-to-leading. The factorizable annihilation amplitudes
$F_{Ha}$, being suppressed only by $M_{\phi,K^*}/M_B$ and almost
imaginary, are the major source of the strong phases in PQCD.
Since the $B\to\phi K^*$ decays are the pure penguin processes
with a weak dependence on the unitarity angle $\phi_3$, these
strong phases determine the relative phases among the helicity
amplitudes $A_0$, $A_\parallel$ and $A_\perp$.

For the $B$ meson wave function, we employ the model \cite{Keum:2001ph},
\begin{eqnarray}
\Phi_{B}(x,b)=N_{B}x^{2}(1-x)^{2}\exp \left[ -\frac{1}{2}
\left( \frac{xM_{B}}{\omega _{B}}\right)^{2}
-\frac{\omega_{B}^{2}b^{2}}{2}\right]
\label{bw} \;,
\end{eqnarray}
where the shape parameter $\omega_{B}=0.4$ GeV has been adopted in
all our previous analyses of exclusive $B$ meson decays. The
normalization constant $N_{B}= 91.784$ GeV is related to the decay
constant $f_{B}=190$ MeV (in the convention $f_{\pi}=130$ MeV). It
is known that there are two $B$ meson wave functions $\Phi_B$ and
$\bar\Phi_B$, which are related to the three-parton $B$ meson wave
functions through a set of equations of motion \cite{GN,BTF,DC,KK}.
Because of the unknown three-parton wave functions, the equations
of motion in fact do not impose any constraint on the functional
form of $\Phi_B$ and $\bar\Phi_B$. Our simple choice of the model
wave functions corresponds to $\Phi_B$ in Eq.~(\ref{bw}) and
${\bar\Phi}_B=0$. This choice is legitimate, since the
contribution from ${\bar\Phi}_B$ is suppressed by a power of
$\bar\Lambda/M_B$ \cite{TLS}, and negligible within the accuracy
of the current formalism.

The $\phi$ and $K^*$ meson distribution amplitudes up to twist 3 are
given by \cite{PB1}
\begin{eqnarray}
\Phi_{\phi }( x) &=&\frac{3f_\phi}{\sqrt{2N_{c}}}x(1-x)\;,
\label{phi2}\\
\Phi_{\phi}^{t}( x )  &=&\frac{f_\phi^T}{2\sqrt{2N_{c}}} \bigg\{
3(1-2x)^{2}+1.68C^{1/2}_4(1-2x)
+0.69\left[ 1+(1-2x)\ln \frac{x}{1-x}\right] \bigg\} \;,
\label{phi3t}\\
\Phi_{\phi}^{s}( x)  &=&\frac{f_\phi^T}{4\sqrt{2N_{c}}} \Big[
3(1-2x)(4.5-11.2x+11.2x^{2})+1.38\ln \frac{x}{1-x}\Big] \;,
\label{phi3s}\\
\Phi_\phi^T(x)&=&\frac{3f_\phi^T}{\sqrt{2N_c}} x(1-x)\Big[1+
0.2C_2^{3/2}(1-2x)\Big]\;,
\label{pwft}\\
\Phi_{\phi}^v(x)&=&\frac{f_{\phi}}{2\sqrt{2N_c}}
\bigg\{\frac{3}{4}[1+(1-2x)^2]+0.24[3(1-2x)^2-1]+0.96C^{1/2}_4(1-2x)
\bigg\}\;,
\label{pwv}\\
\Phi_{\phi}^a(x) &=&\frac{3f_\phi}{4\sqrt{2N_{c}}}
(1-2x)\Big[1+0.93(10x^2-10x+1)\Big]\;,
\label{pwa}\\
\Phi_{K^* }( x) &=&\frac{3f_{K^*}}{\sqrt{2N_{c}}}x(1-x)
\Big[1+0.57(1-2x)+0.07C^{3/2}_2(1-2x)\Big]\;,
\label{pk2}\\
\Phi_{K^*}^{t}( x) &=&\frac{f_{K^*}^T}{2\sqrt{2N_{c}}} \bigg\{
0.3(1-2x)\left[3(1-2x)^2+10(1-2x)-1\right]+1.68C^{1/2}_4(1-2x)
\nonumber \\
&&+0.06(1-2x)^2\left[5(1-2x)^2-3\right]
+0.36\left[ 1-2(1-2x)(1+\ln(1-x))\right] \bigg\} \;,
\label{pk3t}\\
\Phi _{K^*}^s( x)  &=&\frac{f_{K^*}^T}{2\sqrt{2N_{c}}} \bigg\{
3(1-2x)\left[1+0.2(1-2x)+0.6(10x^2-10x+1)\right]
\nonumber \\
& &-0.12x(1-x)+0.36[1-6x-2\ln(1-x)]\bigg\} \;,
\label{pk3s}\\
\Phi_{K^*}^{T}(x)&=&\frac{3f_{K^*}^T}{\sqrt{2N_c}} x(1-x)
\Big[1+0.6(1-2x)+0.04C^{3/2}_2(1-2x)\Big]\;,
\label{pkt}\\
\Phi_{K^*}^v(x)&=&\frac{f_{K^*}}{2\sqrt{2N_c}}
\bigg\{\frac{3}{4}\Big[1+(1-2x)^2+0.44(1-2x)^3\Big]
+0.4C^{1/2}_2(1-2x)
\nonumber \\
&& +0.88C^{1/2}_4(1-2x)+0.48[2x+\ln(1-x)] \bigg\}\;,
\label{pkv}\\
\Phi_{K^*}^a(x) &=&\frac{f_{K^*}}{4\sqrt{2N_{c}}}
\Big\{3(1-2x)\Big[1+0.19(1-2x)+0.81(10x^2-10x+1)\Big]\nonumber \\
&&-1.14x(1-x)+0.48[1-6x-2\ln(1-x)]\Big\}\;,
\label{pka}
\end{eqnarray}
with the Gegenbauer polynomials,
\begin{eqnarray}
C_2^{1/2}(\xi)=\frac{1}{2}(3\xi^2-1)\;,\;\;\;\;
C_4^{1/2}(\xi)=\frac{1}{8}(35 \xi^4 -30 \xi^2 +3)\;,\;\;\;\;
C_2^{3/2}(\xi)=\frac{3}{2}(5\xi^2-1)\;.
\end{eqnarray}

We employ $G_{F}=1.16639\times 10^{-5}$ GeV$^{-2}$, the
Wolfenstein parameters $\lambda =0.2196$, $A=0.819$, and
$R_{b}=0.38$, the unitarity angle $\phi_{3}=90^{o}$, the masses
$M_{B}=5.28$ GeV, $M_{\phi}=1.02$ GeV and $M_{K^*}=0.89$ GeV, the
decay constants $f_{\phi}=237$ MeV, $f_{\phi}^{T}=220$ MeV,
$f_{K^*}=200$ MeV, and $f_{K^*}^{T}=160$ MeV, and the $B_{d}^{0}$
($B^{+}$) meson lifetime $\tau_{B^{0}}=1.55$ ps
($\tau_{B^{+}}=1.65$ ps) \cite{PDG}. We have confirmed that the
above distribution amplitudes and decay constants lead to the
$B\to K^*$ transition form factors \cite{CG} in agreement with
those from light-cone QCD sum rules \cite{ABH}. We have also
confirmed that the averaged values of the running hard scales $t$
defined by Eqs.~(\ref{te12}) and (\ref{et}) in the Appendix
are indeed about $\sqrt{\bar\Lambda M_B}\sim 1.6$ GeV. Note that the
$B\to\phi K^*$ branching ratios are insensitive to the variation
of $\phi_3$. The results for the helicity amplitudes $A_0$,
$A_\parallel$ and $A_\perp$, including their relative phases
$\phi_{\parallel}\equiv Arg(A_{\parallel}/A_{0})$ and
$\phi_{\perp}\equiv Arg(A_{\perp}/A_{0})$, are displayed in Table
I. The contributions to the $B\to\phi K^*$ branching ratios mainly
arise from the longitudinal polarizations $A_{0}$, because of the
relation $|A_{0}|^2\gg |A_{\parallel}|^2\sim |A_{\perp}|^2$, which
is expected from the power counting rules. It is easy to observe
that the ratios $|H_-/H_0|^2$ and $|H_+/H_0|^2$ obtained in PQCD
are close to those in QCDF \cite{CY2}. The annihilation
contributions are the major source of the strong phases, and the
nonfactorizable contributions are the minor one. The values of
$\phi_\parallel$ and $\phi_\perp$ in the rows (I)-(III) of Table
II indicate that the phases from the former are about 4-5 times
those from the latter (but opposite in sign). Without these
sources, we have $\phi_\parallel=\phi_\perp=\pi$. Note that the
relative phases among the different helicity amplitudes can not be
predicted unambiguously in QCDF, due to the arbitrary complex
cutoffs for the evaluation of the nonfactorizable and annihilation
contributions.

We examine the theoretical uncertainty from the variation of the hard
scales $t$, which are defined as the invariant masses of the internal
particles, and required to be higher than the factorization scales
$1/b$, $b$ being the transverse extents of the mesons. This examination
estimates higher-order corrections to the hard
amplitudes, which are the most important theoretical
uncertainty for penguin-dominated $B$ meson decays. The light meson
distribution amplitudes have been determined in QCD sum
rules. The possible 30\% variation of the coefficients of the Gegenbauer
polynomials in these distribution amplitudes lead only to little
changes of our predictions. We consider
the hard scales $t$ located between 0.75-1.25 times the invariant masses
of the internal particles. The predictions for the $B\to\phi K$
branching ratios from the above range are consistent with the data
with uncertainty \cite{CKL}. We then obtain the $B\to\phi K^*$ branching
ratios,
\begin{eqnarray}
B(B_d^0\to\phi K^{*0}) = (14.86^{+4.88}_{-3.36}) \times 10^{-6}\;,
\;\;\;\;
B(B^\pm\to\phi K^{*\pm}) = (15.96^{+5.24}_{-3.61})  \times 10^{-6}\;.
\label{phikb}
\end{eqnarray}
The relative phases $\phi_\parallel$ and $\phi_\perp$, and the
magnitudes $|A_0|^2$, $|A_\parallel|^2$ and $|A_\perp|^2$ of the helicity
amplitudes are quite stable under the variation of the hard
scales $t$. They change within 0.05 $rad.$ and within 0.01, respectively.
There is another minor source of theoretical uncertainty from
the light meson decay constants $f_\phi^{(T)}$ and $f_{K^*}^{(T)}$. If
they reduce by 5\%, the predicted branching ratios will decrease by 10\%.
The CP asymmetries of the $B\to\phi K^*$ modes are, as of $B\to\phi K$,
vanishingly small (less than 2\%).

The above branching ratios are larger than those from QCDF \cite{CY2},
\begin{eqnarray}
B(B_d^0\to\phi K^{*0}) = 8.71 \times 10^{-6}\;,
\;\;\;\;
B(B^\pm\to\phi K^{*\pm}) = 9.30  \times 10^{-6}\;,
\label{phiq}
\end{eqnarray}
due to  the dynamical enhancement of penguin contributions.
We emphasize that the annihilation amplitudes, though not negligible, are
not responsible for the large branching ratios in PQCD, since they are
mainly imaginary. This is understood by comparing the branching ratios in
Table I and in the row (II) of Table II. The nonfactorizable contributions
are not either as shown by the branching ratios in Table I and in the row
(III) of Table II. However, the annihilation contributions, parametrized
as being real, are important in QCDF in order to explain the large
$B\to \phi K$ branching ratios. With the almost real annihilation
contributions, the $B\to\phi K$ branching ratios obtained in QCDF can
increase from $4\times 10^{-6}$ to $7\times 10^{-6}$ \cite{CY1}. The
values quoted in Eq.~(\ref{phiq}) do not include the annihilation
contributions. The current experimental data of
$B(B^0 \to \phi K^{*0})$,
\begin{eqnarray}
{\mathrm CLEO \cite{CLEO}}: & & \hspace{20mm}
(11.5^{+4.5+1.8}_{-3.7-1.7}) \times 10^{-6},
\nonumber\\
{\mathrm BELLE \cite{Belle}}: & & \hspace{20mm}
(15^{+8}_{-6}\pm 3) \times 10^{-6}\;,
\nonumber \\
{\mathrm BABAR \cite{Babar}}: & & \hspace{20mm}
(8.6^{+2.8}_{-2.4}\pm 1.1)\times 10^{-6}\;,
\end{eqnarray}
and those of $B(B^\pm\to \phi K^{*\pm})$,
\begin{eqnarray}
{\mathrm CLEO \cite{CLEO}}: & &\hspace{20mm}
 (10.6^{+6.4+1.8}_{-4.9-1.6})\times 10^{-6},
\nonumber\\
{\mathrm BELLE \cite{Belle}}: & & \hspace{20mm}
<36\times 10^{-6}\;,
\nonumber \\
{\mathrm BABAR \cite{Babar}}: & & \hspace{20mm}
(9.7^{+4.2}_{-3.4}\pm 1.7)\times 10^{-6}\;,
\end{eqnarray}
are not yet precise enough to distinguish the two different approaches.

In this paper we have studied the first observed $B\to VV$ modes,
the $B\to \phi K^*$ decays, using the PQCD formalism. It has been
stressed that two-body heavy meson decays are characterized by a
scale of $O(\bar\Lambda M_B)$ in PQCD, for which penguin
contributions are dynamically enhanced. This enhancement renders
penguin-dominated decay modes acquire branching ratios larger than
those in QCDF, even when the final-state particles are vector
mesons. We have proposed the $B\to\phi K^{(*)}$ decays as the
ideal modes to test the significance of this mechanism. If their
branching ratios are as large as $10 \times 10^{-6}$ ($15\times
10^{-6}$) (independent of the unitarity angle $\phi_3$), dynamical
enhancement will be convincing. We have also emphasized that the
relative importance and the relative strong phases among the
different helicity amplitudes in the $B\to VV$ modes can be
predicted unambiguously in PQCD, which are determined by the power
counting rules and by the annihilation contributions, respectively. These
predictions are insensitive to the variation of the hard scales.
Therefore, the comparison of the results presented
here with future experimental data will provide a stringent
confrontation of the PQCD approach.

\vskip 1.0cm

We thank H.Y. Cheng, K.C. Yang and the members in the PQCD collaboration
for helpful discussions. The work was supported in part by Grant-in Aid
for Special Project Research (Physics of CP Violation), by Grant-in Aid
for Scientific Research from the Ministry of Education, Science and
Culture of Japan. The work of H.N.L. was supported in part by the
National Science Council of R.O.C. under the Grant No.
NSC-90-2112-M-001-077 and by National Center for Theoretical Science of
R.O.C..

\appendix

\section{FACTORIZATION FORMULAS}

In this appendix we present the explicit expressions of the factorizable
and nonfactorizable amplitudes in Eq.~(\ref{dec}). The effective
Hamiltonian for the flavor-changing $b\to s$ transition is given by
\begin{equation}
H_{{\rm eff}}={\frac{G_{F}}{\sqrt{2}}}\sum_{q=u,c}V_{q}\left[ C_{1}(\mu)
O_{1}^{(q)}(\mu )+C_{2}(\mu )O_{2}^{(q)}(\mu )+\sum_{i=3}^{10}C_{i}(\mu)
O_{i}(\mu )\right] \;,  \label{hbk}
\end{equation}
with the CKM matrix elements
$V_{q}=V_{qs}^{*}V_{qb}$ and the operators
\begin{eqnarray}
&&O_{1}^{(q)}=(\bar{s}_{i}q_{j})_{V-A}(\bar{q}_{j}b_{i})_{V-A}\;,\;\;\;\;\;
\;\;\;O_{2}^{(q)}=(\bar{s}_{i}q_{i})_{V-A}(\bar{q}_{j}b_{j})_{V-A}\;,
\nonumber \\
&&O_{3}=(\bar{s}_{i}b_{i})_{V-A}\sum_{q}(\bar{q}_{j}q_{j})_{V-A}\;,\;\;\;
\;O_{4}=(\bar{s}_{i}b_{j})_{V-A}\sum_{q}(\bar{q}_{j}q_{i})_{V-A}\;,
\nonumber \\
&&O_{5}=(\bar{s}_{i}b_{i})_{V-A}\sum_{q}(\bar{q}_{j}q_{j})_{V+A}\;,\;\;\;
\;O_{6}=(\bar{s}_{i}b_{j})_{V-A}\sum_{q}(\bar{q}_{j}q_{i})_{V+A}\;,
\nonumber \\
&&O_{7}=\frac{3}{2}(\bar{s}_{i}b_{i})_{V-A}\sum_{q}e_{q} (\bar{q}%
_{j}q_{j})_{V+A}\;,\;\;O_{8}=\frac{3}{2}(\bar{s}_{i}b_{j})_{V-A}
\sum_{q}e_{q}(\bar{q}_{j}q_{i})_{V+A}\;,  \nonumber \\
&&O_{9}=\frac{3}{2}(\bar{s}_{i}b_{i})_{V-A}\sum_{q}e_{q} (\bar{q}%
_{j}q_{j})_{V-A}\;,\;\;O_{10}=\frac{3}{2}(\bar{s}_{i}b_{j})_{V-A}
\sum_{q}e_{q}(\bar{q}_{j}q_{i})_{V-A}\;,
\end{eqnarray}
$i$ and $j$ being the color indices. Using the unitarity condition, the
CKM matrix elements for the penguin operators $O_{3}$-$O_{10}$ can also
be expressed as $V_{u}+V_{c}=-V_{t}$. The unitarity angle $\phi_{3}$ is
defined via
\begin{equation}
V_{ub}=|V_{ub}|\exp (-i\phi _{3})\;.
\end{equation}
Here we adopt the Wolfenstein parametrization for the CKM matrix upto
$O(\lambda^{3})$,
\begin{eqnarray}
\left(\matrix{V_{ud} & V_{us} & V_{ub} \cr V_{cd} & V_{cs} & V_{cb} \cr
V_{td} & V_{ts} & V_{tb} \cr}\right) =\left(\matrix{ 1 - \lambda^2/2 &
\lambda & A \lambda^3(\rho - i \eta)\cr - \lambda & 1 - \lambda^2/ 2 & A
\lambda^2\cr A \lambda^3(1-\rho-i\eta) & -A \lambda^2 & 1 \cr}\right)\;.
\end{eqnarray}
with the parameters \cite{LEP},
\begin{eqnarray}
\lambda &=& 0.2196 \pm 0.0023\;,  \nonumber \\
A &=& 0.819 \pm 0.035\;,  \nonumber \\
R_b &\equiv&\sqrt{{\rho}^2 + {\eta}^2} = 0.41 \pm 0.07\;.
\end{eqnarray}

The factorizable amplitudes $F_{He}^{(q)}$ and
$F_{Ha}^{(q)}=F_{Ha4}^{(q)}+F_{Ha6}^{(q)}$ are written as
\begin{eqnarray}
F_{Le}^{\left( q\right) } &=&8\pi
C_{F}M_{B}^{2}\int_{0}^{1}dx_{1}dx_{3}\int_{0}^{\infty}
b_{1}db_{1}b_{3}db_{3}\Phi _{B}( x_{1},b_{1})
\nonumber\\
&& \times \left\{ \left [(1+x_{3})\Phi_{K^*}(x_{3}) +
r_{K^*}(1-2x_{3})(\Phi^{t}_{K^*}(x_3)+\Phi^{s}_{K^*}(x_{3}))\right]
\right.
\nonumber\\
&& \times E^{(q)}_{e}(t^{(1)}_{e})
h_{e}(x_{1},x_{3},b_{1},b_{3})
\nonumber\\
&&\left.+2r_{K^{*}}\Phi^{s}_{K^*}(x_3)E^{(q)}_{e}(t^{(2)}_{e})
h_{e}(x_{3},x_{1},b_{3},b_{1}) \right\}\;,
\end{eqnarray}
\begin{eqnarray}
F_{Ne}^{(q) } &=&8\pi
C_{F}M_{B}^{2}\int_{0}^{1}dx_{1}dx_{3}\int_{0}^{\infty}
b_{1}db_{1}b_{3}db_{3}\Phi _{B}(x_{1},b_{1})
\nonumber\\
&& \times r_{\phi} \left\{ [\Phi^{T}_{K^*}(x_3)
+2r_{K^*}\Phi^{v}_{K^*}(x_3)+r_{K^*}x_{3}
(\Phi^{v}_{K^*}(x_3)-\Phi^{a}_{K^*}(x_3))]\right.
\nonumber \\
&& \times E^{(q)}_{e}(t^{(1)}_{e})
h_{e}(x_{1},x_{3},b_{1},b_{3})
\nonumber \\
&&\left.+r_{K^{*}}[\Phi^{v}_{K^*}(x_3)+\Phi^{a}_{K^*}(x_3)]
E^{(q)}_{e}(t^{(2)}_{e})
h_{e}( x_{3},x_{1},b_{3},b_{1})\right\}\;,
\end{eqnarray}
\begin{eqnarray}
F_{Te}^{(q)} &=&16\pi
C_{F}M_{B}^{2}\int_{0}^{1}dx_{1}dx_{3}\int_{0}^{\infty}
b_{1}db_{1}b_{3}db_{3}\Phi _{B}( x_{1},b_{1})
\nonumber\\
&&\times r_{\phi}\left\{[\Phi^{T}_{K^*}(x_3)
+2r_{K^*}\Phi^{a}_{K^*}(x_3)-r_{K^*}x_{3}
(\Phi^{v}_{K^*}(x_3)-\Phi^{a}_{K^*}(x_3))]\right.
\nonumber \\
&& \times E^{(q)}_{e}(t^{(1)}_{e})
h_{e}(x_{1},x_{3},b_{1},b_{3})
\nonumber \\
&&\left.+r_{K^{*}}[\Phi^{v}_{K^*}(x_3)+\Phi^{a}_{K^*}(x_3)]
E^{(q)}_{e}(t^{(2)}_{e})h_{e}( x_{3},x_{1},b_{3},b_{1})\right\}\;,
\end{eqnarray}
\begin{eqnarray}
F_{La4}^{(q) } &=&8\pi
C_{F}M_{B}^{2}\int_{0}^{1}dx_{2}dx_{3}\int_{0}^{\infty}
b_{2}db_{2}b_{3}db_{3}
\nonumber \\
&& \times \left\{ \left[-(1-x_{3})\Phi_{\phi}(x_2)\Phi_{K^*}(x_3)
+2r_{\phi}r_{K^*}\Phi^{s}_{\phi}(x_2)
(x_{3}\Phi^{t}_{K^*}(x_3)+(2-x_{3})\Phi^{s}_{K^*}(x_3))\right] \right.
\nonumber \\
&& \times E_{a4}^{(q)}(t_{a}^{(1)})h_{a}(x_{2},1-x_{3},b_{2},b_{3})
\nonumber \\
&& + \left [ x_{2}\Phi_{\phi}(x_{2})\Phi_{K^*}(x_{3})
+2r_{\phi}r_{K^*}\Phi^{s}_{K^*}(x_{3})
((1-x_{2}) \Phi^{t}_{\phi}(x_2) - (1+x_{2}) \Phi^{s}_{\phi}(x_2))\right]
\nonumber \\
&& \times \left.E_{a4}^{(q)}(t_{a}^{(2)})
h_{a}(1-x_{3},x_{2},b_{3},b_{2}) \right\}\;,
\end{eqnarray}
\begin{eqnarray}
F_{Na4}^{\left( q\right) } &=&-8\pi
C_{F}M_{B}^{2}\int_{0}^{1}dx_{2}dx_{3}\int_{0}^{\infty}
b_{2}db_{2}b_{3}db_{3}
\nonumber \\
&&\times r_{\phi}r_{K^*}\left\{
\left[(2-x_{3})(\Phi^{v}_{\phi}(x_2)\Phi^{v}_{K^*}(x_3)
+\Phi^{a}_{\phi}(x_2)\Phi^{a}_{K^*}(x_3))\right.\right.
\nonumber \\
&& \left.+x_3(\Phi^{v}_{\phi}(x_2)\Phi^{a}_{K^*}(x_3)
+\Phi^{a}_{\phi}(x_2)\Phi^{v}_{K^*}(x_3))\right]
E_{a4}^{(q)}(t_{a}^{(1)})h_{a}(x_{2},1-x_{3},b_{2},b_{3})
\nonumber\\
&& -\left[(1+x_{2})(\Phi^{v}_{\phi}(x_2)\Phi^{v}_{K^*}(x_3)
+\Phi^{a}_{\phi}(x_2)\Phi^{a}_{K^*}(x_3)) \right.
\nonumber \\
&& \left. \left.-(1-x_{2})(\Phi^{v}_{\phi}(x_2)\Phi^{a}_{K^*}(x_3)
+\Phi^{a}_{\phi}(x_2)\Phi^{v}_{K^*}(x_3))\right]
E_{a4}^{(q)}(t_{a}^{(2)})h_{a}(1-x_{3},x_{2},b_{3},b_{2}) \right\}\;,
\end{eqnarray}
\begin{eqnarray}
F_{Ta4}^{\left( q\right) } &=&-16\pi
C_{F}M_{B}^{2}\int_{0}^{1}dx_{2}dx_{3}\int_{0}^{\infty}
b_{2}db_{2}b_{3}db_{3}
\nonumber \\
&&\times r_{\phi}r_{K^*}\left\{
\left[x_{3}(\Phi^{v}_{\phi}(x_2)\Phi^{v}_{K^*}(x_3)
+\Phi^{a}_{\phi}(x_2)\Phi^{a}_{K^*}(x_3))\right.\right.
\nonumber \\
&& \left.+(2-x_3)(\Phi^{v}_{\phi}(x_2)\Phi^{a}_{K^*}(x_3)
+\Phi^{a}_{\phi}(x_2)\Phi^{v}_{K^*}(x_3))\right]
E_{a4}^{(q)}(t_{a}^{(1)})h_{a}(x_{2},1-x_{3},b_{2},b_{3})
\nonumber\\
&& +\left[(1-x_{2})(\Phi^{v}_{\phi}(x_2)\Phi^{v}_{K^*}(x_3)
+\Phi^{a}_{\phi}(x_2)\Phi^{a}_{K^*}(x_3)) \right.
\nonumber \\
&& \left. \left.-(1+x_{2})(\Phi^{v}_{\phi}(x_2)\Phi^{a}_{K^*}(x_3)
+\Phi^{a}_{\phi}(x_2)\Phi^{v}_{K^*}(x_3))\right]
E_{a4}^{(q)}(t_{a}^{(2)})h_{a}(1-x_{3},x_{2},b_{3},b_{2}) \right\}\;,
\end{eqnarray}
\begin{eqnarray}
F_{La6}^{\left( q\right) } &=&16\pi
C_{F}M_{B}^{2}\int_{0}^{1}dx_{2}dx_{3}\int_{0}^{\infty}
b_{2}db_{2}b_{3}db_{3}
\nonumber \\
&& \times \left\{\left[r_{K^*} (1-x_3) \Phi_{\phi}(x_2)
(\Phi^{t}_{K^*}(x_3)+\Phi^{s}_{K^*}(x_3))-2r_{\phi}\Phi^{s}_{\phi}(x_2)
\Phi_{K^*}(x_3) \right] \right.
\nonumber \\
&& \times E_{a6}^{(q)}(t^{(1)}_{a})h_{a}(x_{2},1-x_{3},b_{2},b_{3})
\nonumber\\
&& + \left[r_{\phi} {x_2} (\Phi^{t}_{\phi}(x_2) -
\Phi^{s}_{\phi}(x_2) ) \Phi_{K^*}(x_3)+ 2r_{K^*}
\Phi_{\phi}(x_2)\Phi^{s}_{K^*}(x_3) \right]
\nonumber \\
&& \left.\times E_{a6}^{(q)}(t_{a}^{(2)})h_{a}(1-x_{3},x_{2},b_{3},b_{2})
\right\}\;,
\end{eqnarray}
\begin{eqnarray}
F_{Na6}^{\left( q\right) } &=&16\pi
C_{F}M_{B}^{2}\int_{0}^{1}dx_{2}dx_{3}\int_{0}^{\infty}
b_{2}db_{2}b_{3}db_{3}
\nonumber \\
&& \times\left\{r_{\phi}(\Phi^{v}_{\phi}(x_2)
+\Phi^a_{\phi}(x_{2}))\Phi^{T}_{K^*}(x_3)
E_{a6}^{(q)}(t^{(1)}_{a})h_{a}(x_2,1-x_3,b_2,b_3)\right.
\nonumber \\
&&+ \left. r_{K^*}\Phi^{T}_{\phi}(x_2)(\Phi^{v}_{K^*}(x_3)
-\Phi^a_{\phi}(x_{3}))
E_{a6}^{(q)}(t^{(1)}_{a})h_{a}(1-x_3,x_2,b_3,b_2)\right\}\;,
\end{eqnarray}
\begin{eqnarray}
F_{Ta6}^{( q) }&=&32\pi
C_{F}M_{B}^{2}\int_{0}^{1}dx_{2}dx_{3}\int_{0}^{\infty}
b_{2}db_{2}b_{3}db_{3}
\nonumber \\
&& \times\left\{r_{\phi}(\Phi^{v}_{\phi}(x_2)
+\Phi^a_{\phi}(x_{2}))\Phi^{T}_{K^*}(x_3)
E_{a6}^{(q)}(t^{(1)}_{a})h_{a}(x_2,1-x_3,b_2,b_3)\right.
\nonumber \\
&&+ \left. r_{K^*}\Phi^{T}_{\phi}(x_2)(\Phi^{v}_{K^*}(x_3)
-\Phi^a_{K^*}(x_{3}))
E_{a6}^{(q)}(t^{(2)}_{a})h_{a}(1-x_3,x_2,b_3,b_2)\right\}\;,
\end{eqnarray}
The expression of the factorizable amplitudes $F_{Ha}$ from the tree
operators $O_1$ and $O_2$ are the same as $F_{Ha4}^{(q)}$ but with the
evolution factor $E_{a4}^{(q)}$ replaced by $E_{a1}^{(q)}$.

The factors $E(t)$ contain the evolution from the $W$ boson mass to the
hard scales $t$ in the Wilson coefficients $a(t)$, and from $t$ to the
factorization scale $1/b$ in the Sudakov factors $S(t)$:
\begin{eqnarray}
E_{e}^{(q)}\left( t\right) &=&\alpha _{s}\left( t\right) a_{e}^{(q)}(t)
S_{B}\left( t\right)S_{K^*}\left( t\right)\;,
\nonumber \\
E_{ai}^{(q)}\left( t\right) &=&\alpha _{s}\left( t\right) a_{i}^{(q)}(t)
S_{\phi }(t)S_{K^*}(t) \;.
\label{Eea}
\end{eqnarray}
The Wilson coefficients $a$ in the above formulas are given by
\begin{eqnarray*}
a_{1}^{(q)} &=&C_{2}+\frac{C_{1}}{N_{c}}\;, \\
a_{3}^{(q)} &=&\left( C_{3}+\frac{C_{4}}{N_{c}}\right) +\frac{3}{2}e_{q}
\left(C_{9}+\frac{C_{10}}{N_{c}}\right) \;, \\
a_{4}^{(q)} &=&\left( C_{4}+\frac{C_{3}}{N_{c}}\right) +\frac{3}{2}e_{q}
\left(C_{10}+\frac{C_{9}}{N_{c}}\right) \;, \\
a_{5}^{(q)} &=&\left( C_{5}+\frac{C_{6}}{N_{c}}\right) +\frac{3}{2}e_{q}
\left(C_{7}+\frac{C_{8}}{N_{c}}\right) \;, \\
a_{6}^{(q)} &=&\left( C_{6}+\frac{C_{5}}{N_{c}}\right) +\frac{3}{2}e_{q}
\left(C_{8}+\frac{C_{7}}{N_{c}}\right) \;, \\
a_{e}^{(q)} &=&a_{3}^{(q)}+a_{4}^{(q)}+a_{5}^{(q)}\;.
\end{eqnarray*}
$k_T$ resummation of large logarithmic corrections to the $B$, $\phi$
and $K^*$ meson distribution amplitudes lead to the exponentials
$S_{B}$, $S_{\phi}$ and $S_{K^*}$, respectively,
\begin{eqnarray}
S_{B}(t)&=&\exp\left[-s(x_{1}P_{1}^{+},b_{1})
-2\int_{1/b_{1}}^{t}\frac{d{\bar{\mu}}} {\bar{\mu}}
\gamma (\alpha _{s}({\bar{\mu}}^2))\right]\;,
\nonumber \\
S_{\phi }(t)&=&\exp\left[-s(x_{2}P_{2}^{+},b_{2})
-s((1-x_{2})P_{2}^{+},b_{2})
-2\int_{1/b_{2}}^{t}\frac{d{\bar{\mu}}}{\bar{\mu}}
\gamma (\alpha _{s}({\bar{\mu}}^2))\right]\;,
\nonumber \\
S_{K^*}(t)&=&\exp\left[-s(x_{3}P_{3}^{-},b_{3})
-s((1-x_{3})P_{3}^{-},b_{3})
-2\int_{1/b_{3}}^{t}\frac{d{\bar{\mu}}}{\bar{\mu}}
\gamma (\alpha_{s}({\bar{\mu}}^2))\right]\;,
\label{sbk}
\end{eqnarray}
with the quark anomalous dimension $\gamma=-\alpha_s/\pi$.
The variables $b_{1}$, $b_{2}$, and $b_{3}$, conjugate to the parton
transverse momenta $k_{1T}$, $k_{2T}$, and $k_{3T}$, represent the
transverse extents of the $B$, $\phi$ and $K^*$ mesons, respectively.
The expression for the exponent $s$ is referred to \cite{BS,LS}.
The above Sudakov exponentials decrease fast in the large $b$ region
\cite{TLS,WY}, such that the $B\to\phi K^*$ hard amplitudes remain
sufficiently perturbative in the end-point region.

The hard functions $h$'s are
\begin{eqnarray}
h_{e}(x_{1},x_{3},b_{1},b_{3}) &=&K_{0}\left( \sqrt{x_{1}x_{3}}
M_{B}b_{1}\right)S_t(x_3)
\nonumber \\
&&\times \left[ \theta (b_{1}-b_{3})K_{0}\left( \sqrt{x_{3}}
M_{B}b_{1}\right) I_{0}\left( \sqrt{x_{3}}M_{B}b_{3}\right) \right.
\nonumber \\
&&\left. +\theta (b_{3}-b_{1})K_{0}\left( \sqrt{x_{3}}M_{B}b_{3}\right)
I_{0}\left( \sqrt{x_{3}}M_{B}b_{1}\right) \right] \;,
\label{he} \\
h_{a}(x_{2},x_{3},b_{2},b_{3}) &=&\left( \frac{i\pi }{2}\right)^{2}
H_{0}^{(1)}\left( \sqrt{x_{2}x_{3}}M_{B}b_{2}\right)S_t(x_3)
\nonumber \\
&&\times \left[ \theta (b_{2}-b_{3})H_{0}^{(1)}\left( \sqrt{x_{3}}
M_{B}b_{2}\right) J_{0}\left( \sqrt{x_{3}}M_{B}b_{3}\right) \right.
\nonumber \\
&&\left. +\theta (b_{3}-b_{2})H_{0}^{(1)}\left( \sqrt{x_{3}}
M_{B}b_{3}\right) J_{0}\left( \sqrt{x_{3}}M_{B}b_{2}\right) \right] \;.
\label{ha}
\end{eqnarray}
We have proposed the parametrization for the evolution function
$S_{t}(x)$ from threshold resummation \cite{TLS,L5},
\begin{eqnarray}
S_t(x)=\frac{2^{1+2c}\Gamma(3/2+c)}{\sqrt{\pi}\Gamma(1+c)} [x(1-x)]^c\;.
\label{str}
\end{eqnarray}
where the parameter $c$ is chosen as $c=0.4$ for the $B\to\phi K^*$
decays. This factor modifies the end-point behavior of the meson
distribution amplitudes, rendering them vanish faster at $x\to 0$.
Threshold resummation for nonfactorizable diagrams is weaker and
negligible. $K_0, I_0, H_0$ and $J_0$ are the Bessel functions.

The hard scales $t$ are chosen as the maxima of the virtualities of the
internal particles involved in the hard amplitudes, including $1/b_{i}$:
\begin{eqnarray}
t_{e}^{(1)} &=&{\rm max}(\sqrt{x_{3}}M_{B},1/b_{1},1/b_{3})\;,
\nonumber \\
t_{e}^{(2)} &=&{\rm max}(\sqrt{x_{1}}M_{B},1/b_{1},1/b_{3})\;,
\label{te12} \\
t_{a}^{(1)} &=&{\rm max}(\sqrt{1-x_{3}}M_{B},1/b_{2},1/b_{3})\;,
\nonumber\\
t_{a}^{(2)} &=&{\rm max}(\sqrt{x_{2}}M_{B},1/b_{2},1/b_{3})\;.
\label{et}
\end{eqnarray}
When the PQCD formalism is extended to $O(\alpha_s^2)$, the hard scales
can be determined more precisely and the scale independence of our
predictions will be improved. Before this calculation is carried out,
we consider the variation of, for example, $t_e$ in the following range,
\begin{eqnarray}
{\rm max}(0.75\sqrt{x_{3}}M_{B},1/b_{1},1/b_{3})<t_{e}^{(1)}<
{\rm max}(1.25\sqrt{x_{3}}M_{B},1/b_{1},1/b_{3})\;,
\nonumber \\
{\rm max}(0.75\sqrt{x_{1}}M_{B},1/b_{1},1/b_{3})<
t_{e}^{(2)} <{\rm max}(1.25\sqrt{x_{1}}M_{B},1/b_{1},1/b_{3})\;,
\end{eqnarray}
in order to estimate the $O(\alpha_s^2)$ corrections. The range for
$t_a$ is chosen in a similar way.

The nonfactorizable amplitudes ${\cal M}_{He}^{( q) }=
{\cal M}_{He3}^{\left( q\right) }+{\cal M}_{He4}^{\left( q\right) }
+{\cal M}_{He5}^{\left( q\right) }
+{\cal M}_{He6}^{\left( q\right) }$ and ${\cal M}_{Ha}^{( q) }=
{\cal M}_{Ha3}^{\left( q\right) }+{\cal M}_{Ha5}^{\left( q\right) }$,
depending on kinematic variables of all the three mesons \cite{WYL}, are
written as
\begin{eqnarray}
{\cal M}_{Le3}^{( q) } &=&16\pi C_{F}M_{B}^{2}\sqrt{2N_{c}}%
\int_{0}^{1}d[x]\int_{0}^{\infty }b_{1}db_{1}b_{2}db_{2}
\Phi _{B}(x_{1},b_{1})
\nonumber \\
&&\times \left\{\Phi _{\phi }(x_{2})
\left[ -(x_{2}+x_{3}) \Phi _{K^{*}} ( x_{3} ) +r_{K^*} x_3
(\Phi^{t}_{K^*}(x_3)+\Phi^{s}_{K^*}(x_3)) \right] \right.
\nonumber\\
&& \times E_{e3}^{(q)\prime}(t_d^{(1)})h_d^{(1)}(x_1,x_2,x_3,b_1,b_2)
\nonumber \\&&
+\Phi _{\phi }(x_{2}) \left[ (1-x_{2}) \Phi_{K^{*}} ( x_{3} )
+r_{K^*}x_3(\Phi^{t}_{K^*}(x_3)-\Phi^{s}_{K^*}(x_3)) \right]
\nonumber \\
&&\left. \times E_{e3}^{(q)\prime}(t^{(2)}_d)
h_d^{(2)}(x_1,x_2,x_3,b_1,b_2) \right\}\;,
\end{eqnarray}
\begin{eqnarray}
{\cal M}_{Ne3}^{( q) } &=&16\pi C_{F}M_{B}^{2}\sqrt{2N_{c}}
\int_{0}^{1}d[x]\int_{0}^{\infty }b_{1}db_{1}b_{2}db_{2}
\Phi _{B}(x_{1},b_{1})
\nonumber \\
&&\times r_{\phi}\left\{ \left[x_2(\Phi^{v}_{\phi}(x_2)
+\Phi^a_{\phi}(x_2))\Phi^{T}_{K^*}(x_3)\right.\right.
\nonumber \\
&& \left. -2r_{K^*}(x_2+x_3)
(\Phi^{v}_{\phi}(x_2) \Phi^{v}_{K^*}(x_3)+ \Phi^{a}_{\phi}(x_2)
\Phi^{a}_{K^*}(x_3) ) \right]
\nonumber\\
&& \times E_{e3}^{(q)\prime}(t^{(1)}_d)h_d^{(1)}(x_1,x_2,x_3,b_1,b_2)
\nonumber \\
&& + (1-x_2)(\Phi^{v}_{\phi}(x_2)+\Phi^{a}_{\phi}(x_2) )
\Phi^{T}_{K^*}(x_3)
\nonumber \\
&&\left. \times E_{e3}^{(q)\prime}(t^{(2)}_d)
h_d^{(2)}(x_1,x_2,x_3,b_1,b_2) \right\}\;,
\end{eqnarray}
\begin{eqnarray}
{\cal M}_{Te3}^{\left( q\right) } &=&32\pi C_{F}M_{B}^{2}\sqrt{2N_{c}}
\int_{0}^{1}d[x]\int_{0}^{\infty }b_{1}db_{1}b_{2}db_{2}
\Phi _{B}(x_{1},b_{1})
\nonumber \\
&&\times  r_{\phi}\left\{ \left[
x_2(\Phi^{v}_{\phi}(x_2)+\Phi^a_{\phi}(x_2))\Phi^{T}_{K^*}(x_3)
\right.\right.
\nonumber \\ &&
\left. -2r_{K^*}(x_2+x_3)(\Phi^{v}_{\phi}(x_2)
\Phi^{a}_{K^*}(x_3)+ \Phi^{a}_{\phi}(x_2)
\Phi^{v}_{K^*}(x_3) ) \right]
\nonumber\\
&& \times E_{e3}^{(q)\prime}(t^{(1)}_d)h_d^{(1)}(x_1,x_2,x_3,b_1,b_2)
\nonumber \\
&& + (1-x_2)(\Phi^{v}_{\phi}(x_2)+\Phi^{a}_{\phi}(x_2) )
\Phi^{T}_{K^*}(x_3)
\nonumber \\
&&\left. \times E_{e3}^{(q)\prime}(t^{(2)}_d)
h_d^{(2)}(x_1,x_2,x_3,b_1,b_2) \right\}\;,
\end{eqnarray}
\begin{eqnarray}
{\cal M}_{Le5}^{(q) } &=&16\pi C_{F}M_{B}^{2}\sqrt{2N_{c}}
\int_{0}^{1}d[x]\int_{0}^{\infty }b_{1}db_{1}b_{2}db_{2}
\Phi_{B}(x_{1},b_{1})
\nonumber \\
&&\times r_{\phi} \left \{
\left[ -x_2 ( \Phi^{t}_{\phi}(x_2)-\Phi^{s}_{\phi}(x_2))
\Phi_{K^*}(x_3) \right.\right.
\nonumber \\
&&  +r_{K^*}x_2(\Phi^{t}_{\phi}(x_2)
-\Phi^{s}_{\phi}(x_2))(\Phi^{t}_{K^*}(x_3)-\Phi^{s}_{K^*}(x_3))
\nonumber\\
&& \left. +r_{K^*}x_{3}(\Phi^{t}_{\phi}(x_2)+\Phi^{s}_{\phi}(x_2))
(\Phi^{t}_{K^*}(x_3)+\Phi^{s}_{K^*}(x_3)) \right]
\nonumber \\
&& \times E^{(q)\prime}_{e5}(t^{(1)}_{d})
h_{d}^{(1)}(x_{1},x_{2},x_{3},b_{1},b_{2})
\nonumber \\
&& +\left[-(1-x_2)(\Phi^{t}_{\phi}(x_2)+\Phi^{s}_{\phi}(x_2))
\Phi_{K^*}(x_3)\right.
\nonumber \\
&& +r_{K^*}(1-x_2)(\Phi^{t}_{\phi}(x_2)
+\Phi^{s}_{\phi}(x_2))(\Phi^{t}_{K^*}(x_3)-\Phi^{s}_{K^*}(x_3))
\nonumber \\
&& \left. + r_{K^*} x_3(\Phi^{t}_{\phi}(x_2)
-\Phi^{s}_{\phi}(x_2))(\Phi^{t}_{K^*}(x_3)+\Phi^{s}_{K^*}(x_3))
\right]
\nonumber \\
&& \left. \times E_{e5}^{(q)\prime}(t^{(2)}_d)h_{d}^{(2)}(
x_{1},x_{2},x_{3},b_{1},b_{2}) \right\}\;,
\end{eqnarray}
\begin{eqnarray}
{\cal M}_{Ne5}^{(q) } &=&-16\pi C_{F}M_{B}^{2}\sqrt{2N_{c}}
\int_{0}^{1}d[x]\int_{0}^{\infty }b_{1}db_{1}b_{2}db_{2}
\Phi _{B}(x_{1},b_{1}) \nonumber \\ &&\times  r_{K^*}x_{3}
\Phi^{T}_{\phi}(x_2) (\Phi^{v}_{K^*}(x_3)-\Phi^{a}_{K^*}(x_3))
\nonumber \\
&&\times \left \{
E^{(q)\prime}_{e5}(t^{(1)}_{d})h_{d}^{(1)}(x_{1},x_{2},x_{3},b_{1},b_{2})
+ E^{(q)\prime}_{e5}(t^{(2)}_{d})h_{d}^{(2)}(x_{1},x_{2},x_{3},b_{1},b_{2})
\right\}\;,
\end{eqnarray}
\begin{eqnarray}
{\cal M}_{Te5}^{\left( q\right) }= 2{\cal M}_{Ne5}^{(q) }\;,
\end{eqnarray}
\begin{eqnarray}
{\cal M}_{Le6}^{(q) } &=&-16\pi C_{F}M_{B}^{2}\sqrt{2N_{c}}
\int_{0}^{1}d[x]\int_{0}^{\infty }b_{1}db_{1}b_{2}db_{2}
\Phi _{B}(x_{1},b_{1})
\nonumber \\
&&\times \Phi_{\phi}(x_2)\left\{ \left[
x_2\Phi_{K^*}(x_3)+r_{K^*}x_{3}(\Phi^{t}_{K^*}(x_3)-\Phi^{s}_{K^*}(x_3))
\right] \right.
\nonumber \\
&& \times E^{(q)\prime}_{e5}(t^{(1)}_{d})
h_{d}^{(1)}(x_{1},x_{2},x_{3},b_{1},b_{2})
\nonumber \\
&&  + \left[-(1-x_2+x_3)\Phi_{K^*}(x_3) + r_{K^*}x_3
(\Phi^{t}_{K^*}(x_3)+\Phi^{s}_{K^*}(x_3))\right]
\nonumber\\
&& \left.\times E^{(q)\prime}_{e5}(t^{(2)}_{d})
h_{d}^{(2)}(x_{1},x_{2},x_{3},b_{1},b_{2})\right\}\;,
\end{eqnarray}
\begin{eqnarray}
{\cal M}_{Ne6}^{(q) } &=&-16\pi C_{F}M_{B}^{2}\sqrt{2N_{c}}
\int_{0}^{1}d[x]\int_{0}^{\infty }b_{1}db_{1}b_{2}db_{2}
\Phi _{B}(x_{1},b_{1})
\nonumber \\
&&\times r_{\phi} \left\{x_2(\Phi^{v}_{\phi}(x_2)-\Phi^{a}_{\phi}(x_2))
\Phi^{T}_{K^*}(x_3)E^{(q)\prime}_{e5}(t^{(1)}_{d})
h_{d}^{(1)}(x_{1},x_{2},x_{3},b_{1},b_{2})\right.
\nonumber \\
&& +  \left[(1-x_2)(\Phi^{v}_{\phi}(x_2)-\Phi^{a}_{\phi}(x_2))
\Phi^{T}_{K^*}(x_3)\right.
\nonumber\\
&&\left. -2r_{K^*}(1-x_2+x_3)(\Phi^{v}_{\phi}(x_2)
\Phi^{v}_{K^*}(x_3)-\Phi^{a}_{\phi}(x_2)\Phi^{a}_{K^*}(x_3)) \right]
\nonumber \\
&& \left. \times E^{(q)\prime}_{e5}(t^{(2)}_{d})
h_{d}^{(2)}(x_{1},x_{2},x_{3},b_{1},b_{2})\right\}\;,
\end{eqnarray}
\begin{eqnarray}
{\cal M}_{Te6}^{(q) } &=&-32\pi C_{F}M_{B}^{2}\sqrt{2N_{c}}
\int_{0}^{1}d[x]\int_{0}^{\infty }b_{1}db_{1}b_{2}db_{2}
\Phi _{B}(x_{1},b_{1})
\nonumber \\
&&\times r_{\phi} \left\{x_2(\Phi^{v}_{\phi}(x_2)-\Phi^{a}_{\phi}(x_2))
\Phi^{T}_{K^*}(x_3)E^{(q)\prime}_{e5}(t^{(1)}_{d})
h_{d}^{(1)}(x_{1},x_{2},x_{3},b_{1},b_{2})\right.
\nonumber \\
&& + \left[(1-x_2)(\Phi^{v}_{\phi}(x_2)-\Phi^{a}_{\phi}(x_2))
\Phi^{T}_{K^*}(x_3)\right.
\nonumber\\
&&\left. -2r_{K^*}(1-x_2+x_3)(\Phi^{v}_{\phi}(x_2)\Phi^{a}_{K^*}(x_3)
-\Phi^{a}_{\phi}(x_2)\Phi^{v}_{K^*}(x_3)) \right]
\nonumber \\
&& \left. \times E^{(q)\prime}_{e5}(t^{(2)}_{d})
h_{d}^{(2)}(x_{1},x_{2},x_{3},b_{1},b_{2})\right\}\;,
\end{eqnarray}
\begin{eqnarray}
{\cal M}_{La3}^{( q) } &=&16\pi C_{F}M_{B}^{2}\sqrt{2N_{c}}
\int_{0}^{1}d[x]\int_{0}^{\infty }b_{1}db_{1}b_{2}db_{2}
\Phi _{B}(x_{1},b_{1})
\nonumber \\
&&\times \left\{ \left[ (1-x_3)
\Phi_{\phi}(x_2)\Phi_{K^*}(x_3)+r_{\phi}r_{K^*}
\left((1+x_2-x_3)(\Phi^{t}_{\phi}(x_2)\Phi^{t}_{K^*}(x_3) \right.
\right. \right.
\nonumber\\
&& \left. \left. \left. -\Phi^{s}_{\phi}(x_2)\Phi^{s}_{K^*}(x_3))
-(1-x_2-x_3)(\Phi^{t}_{\phi}(x_2)\Phi^{s}_{K^*}(x_3)
-\Phi^{s}_{\phi}(x_2)\Phi^{t}_{K^*}(x_3) ) \right) \right] \right.
\nonumber \\
&& \times E^{(q)\prime}_{a3}(t^{(1)}_{f})
h_{f}^{(1)}(x_{1},x_{2},x_{3},b_{1},b_{2})
\nonumber \\
&& - \left[x_2\Phi_{\phi}(x_2)\Phi_{K^*}(x_3)
-2r_{\phi}r_{K^*}(\Phi^{t}_{\phi}(x_2)\Phi^{t}_{K^*}(x_3)
+\Phi^{s}_{\phi}(x_2)\Phi^{s}_{K^*}(x_3)) \right.
\nonumber \\
&&+r_{\phi}r_{K^*}(1+x_2-x_3)(\Phi^{t}_{\phi}(x_2)\Phi^{t}_{K^*}(x_3)
-\Phi^{s}_{\phi}(x_2)\Phi^{s}_{K^*}(x_3))
\nonumber \\
&& \left.+r_{\phi}r_{K^*}
(1-x_2-x_3)(\Phi^{t}_{\phi}(x_2)\Phi^{s}_{K^*}(x_3)
-\Phi^{s}_{\phi}(x_2)\Phi^{t}_{K^*}(x_3))\right]
\nonumber\\
&& \left.\times E^{(q)\prime}_{a3}(t^{(2)}_{d})
h_{f}^{(2)}(x_{1},x_{2},x_{3},b_{1},b_{2}) \right\}\;,
\end{eqnarray}
\begin{eqnarray}
{\cal M}_{Na3}^{(q) } &=&-32\pi C_{F}M_{B}^{2}\sqrt{2N_{c}}
\int_{0}^{1}d[x]\int_{0}^{\infty }b_{1}db_{1}b_{2}db_{2}
\Phi _{B}(x_{1},b_{1})
\nonumber \\
&&\times r_{\phi}r_{K^*} \left[\Phi^{v}_{\phi}(x_2)\Phi^{v}_{K^*}(x_3)
+\Phi^{a}_{\phi}(x_2)\Phi^{a}_{K^*}(x_3)\right]
E^{(q)\prime}_{a3}(t^{(2)}_{f})
h_{f}^{(2)}(x_{1},x_{2},x_{3},b_{1},b_{2})\;,
\end{eqnarray}
\begin{eqnarray}
{\cal M}_{Ta3}^{(q) } &=&-64\pi C_{F}M_{B}^{2}\sqrt{2N_{c}}
\int_{0}^{1}d[x]\int_{0}^{\infty }b_{1}db_{1}b_{2}db_{2}
\Phi _{B}(x_{1},b_{1})
\nonumber \\
&&\times r_{\phi}r_{K^*} \left[\Phi^{v}_{\phi}(x_2)\Phi^{a}_{K^*}(x_3)
+\Phi^{a}_{\phi}(x_2)\Phi^{v}_{K^*}(x_3)\right]
E^{(q)\prime}_{a3}(t^{(2)}_{f})
h_{f}^{(2)}(x_{1},x_{2},x_{3},b_{1},b_{2})\;,
\end{eqnarray}
\begin{eqnarray}
{\cal M}_{La5}^{(q) } &=&16\pi C_{F}M_{B}^{2}\sqrt{2N_{c}}
\int_{0}^{1}d[x]\int_{0}^{\infty }b_{1}db_{1}b_{2}db_{2}
\Phi _{B}(x_{1},b_{1})
\nonumber \\
&&\times \left\{ \left[ r_{K^*}(1-x_3)
\Phi_{\phi}(x_2)(\Phi^{t}_{K^*}(x_3)-
\Phi^{s}_{K^*}(x_3))-r_{\phi}x_{2}(\Phi^{t}_{\phi}(x_2)
+\Phi^{s}_{\phi}(x_2))\Phi_{K^*}(x_3)\right] \right.
\nonumber \\
&& \times E^{(q)\prime}_{a5}(t^{(1)}_{f})
h_{f}^{(1)}(x_{1},x_{2},x_{3},b_{1},b_{2})
\nonumber \\
&&  + \left[-r_{\phi}(2-x_2)(\Phi^{t}_{\phi}(x_2)+\Phi^{s}_{\phi}(x_2))
\Phi_{K^*}(x_3) \right.
\nonumber \\
&&\left.\left.+r_{K^*}(1+x_3)\Phi_{\phi}(x_2)
(\Phi^{t}_{K^*}(x_3)-\Phi^{s}_{K^*}(x_3))\right]
E^{(q)\prime}_{a5}(t^{(2)}_{f})h_{f}^{(2)}(x_{1},x_{2},x_{3},b_{1},b_{2})
\right\}\;,
\end{eqnarray}
\begin{eqnarray}
{\cal M}_{Na5}^{(q) } &=&16\pi C_{F}M_{B}^{2}\sqrt{2N_{c}}
\int_{0}^{1}d[x]\int_{0}^{\infty }b_{1}db_{1}b_{2}db_{2}
\Phi _{B}(x_{1},b_{1})
\nonumber \\
&&\times \left\{\left[r_{\phi}x_2(\Phi^{v}_{\phi}(x_2)
+\Phi^{a}_{\phi}(x_2))\Phi^T_{K^*}(x_3)-
r_{K^*} (1-x_3) \Phi^T_{\phi}(x_2)
(\Phi^{v}_{K^*}(x_3)-\Phi^{a}_{K^*}(x_3))\right] \right.
\nonumber\\
&& \times E^{(q)\prime}_{a5}(t^{(1)}_{f})
h_{d}^{(1)}(x_{1},x_{2},x_{3},b_{1},b_{2})
\nonumber \\
&& +\left[r_{\phi}(2-x_2)(\Phi^{v}_{\phi}(x_2)
+\Phi^{a}_{\phi}(x_2))\Phi^{T}_{K^*}(x_3)
-r_{K^*}(1+x_3)\Phi^{T}_{\phi}(x_2)
(\Phi^{v}_{K^*}(x_3)-\Phi^{a}_{K^*}(x_3)) \right]
\nonumber \\
&& \times \left. E^{(q)\prime}_{a5}(t^{(2)}_{f})
h_{f}^{(2)}(x_{1},x_{2},x_{3},b_{1},b_{2})\right\}\;,
\end{eqnarray}
\begin{eqnarray}
{\cal M}_{Ta5}^{(q) }= 2 {\cal M}_{Na5}^{(q) }\;.
\end{eqnarray}
The expressions of the nonfactorizable amplitudes ${\cal M}_{Ha}$ and
${\cal M}_{He4}$ are the same as ${\cal M}_{Ha3}^{(q)}$ and
${\cal M}_{He3}^{(q)}$ but with the evolution factors $E_{a3}^{(q)\prime}$
and $E_{e3}^{(q)\prime}$ replaced by $E_{a1}^{(q)\prime}$ and
$E_{e4}^{(q)\prime}$, respectively.

The evolution factors are given by
\begin{eqnarray}
E_{ei}^{\left( q\right) \prime }\left( t\right) &=&\alpha _{s}\left(
t\right) a_{i}^{\left( q\right)\prime}(t)S\left( t\right)|_{b_{3}=b_{1}}\;,
 \nonumber \\
E_{ai}^{\left( q\right) \prime }\left( t\right)
&=&\alpha _{s}\left( t\right)
a_{i}^{\left( q\right)\prime}(t)S\left( t\right)|_{b_{3}=b_{2}}\;,
\label{Eeaprim}
\end{eqnarray}
with the Sudakov factor $S=S_{B}S_{\phi }S_{K^*}$.
The Wilson coefficients $a$ appearing in the above formulas are
\begin{eqnarray*}
a_{1}^{\prime }&=&\frac{C_{1}}{N_{c}}\;, \\
a_{3}^{(q)\prime} &=&\frac{1}{N_{c}}
\left(C_{3}+\frac{3}{2}e_{q}C_{9}\right)\;,\\
a_{4}^{(q)\prime}&=&\frac{1}{N_{c}}
\left( C_{4}+\frac{3}{2}e_{q}C_{10}\right)\;,\\
a_{5}^{(q)\prime}&=&\frac{1}{N_{c}}
\left( C_{5}+\frac{3}{2}e_{q}C_{7}\right) \;,\\
a_{6}^{(q)\prime} &=&\frac{1}{N_{c}}
\left(C_{6}+\frac{3}{2}e_{q}C_{8}\right) \;.
\end{eqnarray*}

The hard functions $h^{(j)}$, $j=1$ and 2, are written as
\begin{eqnarray}
h_{d}^{(j)} &=&\left[ \theta (b_{1}-b_{2})K_{0}\left( DM_{B}b_{1}\right)
I_{0}\left( DM_{B}b_{2}\right) \right.
\nonumber \\
&&\quad \left. +\theta (b_{2}-b_{1})K_{0}\left( DM_{B}b_{2}\right)
I_{0}\left( DM_{B}b_{1}\right) \right]  \nonumber \\
&&\times K_{0}(D_{j}M_{B}b_{2})\;,\;\;\;\;\;\;\;\;\;\;\;\;\;\;\;\;\;\;\;
\mbox{for $D^2_{j} \geq 0$}\;,
\nonumber \\
&&\times \frac{i\pi }{2}H_{0}^{(1)}
\left(\sqrt{|D_{j}^{2}|}M_{B}b_{2}\right)\;,\;\;\;\;
\mbox{for $D^2_{j} \leq 0$}\;,
\label{hjd} \\
h_{f}^{(j)} &=&\frac{i\pi }{2}\left[ \theta (b_{1}-b_{2})H_{0}^{(1)}
\left(FM_{B}b_{1}\right) J_{0}\left( FM_{B}b_{2}\right) \right.
\nonumber \\
&&\quad \left. +\theta (b_{2}-b_{1})H_{0}^{(1)}\left( FM_{B}b_{2}\right)
J_{0}\left( FM_{B}b_{1}\right) \right]
\nonumber \\
&&\times K_{0}(F_{j}M_{B}b_{1})\;,\;\;\;\;\;\;\;\;\;\;\;\;\;\;\;\;\;\;\;
\mbox{for $F^2_{j} \geq 0$}\;,
\nonumber \\
&&\times \frac{i\pi }{2}H_{0}^{(1)}
\left(\sqrt{|F_{j}^{2}|}M_{B}b_{1}\right)\;,\;\;\;\;
\mbox{for $F^2_{j} \leq 0$}\;,
\label{hjf}
\end{eqnarray}
with the variables,
\begin{eqnarray}
D^{2} &=& x_{1}x_{3}\;,
\nonumber \\
D_{1}^{2} &=&(x_{1}-x_{2})x_{3}\;,
\nonumber \\
D_{2}^{2} &=&-(1-x_{1}-x_{2})x_{3}\;, \\
F^{2} &=& x_{2} \left( 1-x_{3}\right) \;,
\nonumber \\
F_{1}^{2} &=&(x_{1}-x_{2}) \left( 1-x_{3}\right)\;,
\nonumber \\
F_{2}^{2} &=&x_{1}+x_{2}+\left(1- x_{1}-x_{2}\right)
\left( 1-x_{3}\right)\;.
\label{DF}
\end{eqnarray}
The hard scales $t^{(j)}$ are chosen as
\begin{eqnarray}
t_{d}^{(1)} &=&{\rm max}\left(DM_{B},\sqrt{|D_{1}^{2}|}M_{B},1/b_{1},
1/b_{2}\right)\;,
\nonumber \\
t_{d}^{(2)} &=&{\rm max}\left(DM_{B},\sqrt{|D_{2}^{2}|}M_{B},1/b_{1},
1/b_{2}\right)\;,
\nonumber \\
t_{f}^{(1)} &=&{\rm max}\left(FM_{B},\sqrt{|F_{1}^{2}|}M_{B},1/b_{1},
1/b_{2}\right)\;,
\nonumber \\
t_{f}^{(2)} &=&{\rm max}\left(FM_{B},\sqrt{|F_{2}^{2}|}M_{B},1/b_{1},
1/b_{2}\right)\;.
\end{eqnarray}

\newpage 
\begin{table}[htbp]
\caption{Helicity amplitudes and relative phases.} \label{table1}
\begin{center}
\begin{tabular}{ccccccc}
\hline
 Mode& $BR(10^{-6})$ &$ |A_{0}|^{2}$ & $ |A_{\parallel}|^{2}$
 & $ |A_{\perp}|^{2}$
 & $\phi_{\parallel}(rad.)$ & $\phi_{\perp}(rad.)$ \\ \hline
 $\phi K^{*0}$& $14.86$ &  $0.750$ & $0.135$ & $0.115$ & $2.55$ & $2.54$ \\
  \hline
  $\phi K^{*+}$ & $15.96$ &$0.748$ & $0.133$ & $0.111$ & $2.55$ & $2.54$ \\
   \hline
\end{tabular}
\end{center}
\end{table}
\begin{table}[htbp]
\caption{Helicity amplitudes and relative phases:
(I) without annihilation and nonfactorizable contributions,
(II) without annihilation contributions, and (III) without nonfactorizable
contributions.}\label{table2}
\begin{center}
\begin{tabular}{ccccccc}
\hline
Mode& $BR(10^{-6})$ & $ |A_{0}|^{2}$ & $ |A_{\parallel}|^{2}$
& $ |A_{\perp}|^{2}$ & $\phi_{\parallel}(rad.)$ & $\phi_{\perp}(rad.)$
\\ \hline
$\phi K^{*0}$(I) & $14.48$ & $0.923$  & $0.040$ & $0.035$ &
$\pi$ & $\pi$\\
\hspace{0.7cm}(II)& $13.25$ &  $0.860$ & $0.072$ & $0.063$ & $3.30$
& $3.33$ \\
\hspace{0.7cm}(III)  & $16.80$ &  $0.833$ & $0.089$ & $0.078$ &
$2.37$ & $2.34$ \\ \hline
$\phi K^{*+}$(I)  & $15.45$ &  $0.923$ & $0.040$ & $0.035$  &
$\pi$  & $\pi$  \\
\hspace{0.7cm}(II) & $14.17$ &$0.860$ & $0.072$ & $0.063$ & $3.30$ & $3.33$
\\
\hspace{0.7cm}(III)  & $17.98$ &$0.830$ & $0.094$ & $0.075$ & $2.37$ &
 $2.34$ \\\hline
\end{tabular}
\end{center}
\end{table}


\begin{references}

\bibitem{Keum:2001ph} Y.Y. Keum, H-n. Li, and A.I. Sanda,
Phys Lett. B {\bf 504}, 6 (2001);
Phys. Rev. D {\bf 63}, 054008 (2001).
\bibitem{KL} Y.Y. Keum and H-n. Li, Phys. Rev.
{\bf D63}, 074006 (2001).
\bibitem{LUY} C. D. L\"{u}, K. Ukai, and M. Z. Yang, Phys. Rev. D {\bf 63},
074009 (2001).
\bibitem{BBNS}  M. Beneke, G. Buchalla, M. Neubert, and C.T. Sachrajda,
Phys. Rev. Lett. {\bf 83}, 1914 (1999);
Nucl. Phys. {\bf B591}, 313 (2000).
\bibitem{BBNS2}  M. Beneke, G. Buchalla, M. Neubert, and C.T. Sachrajda,
Nucl. Phys. {\bf B606}, 245 (2001).
\bibitem{L6} H-n. Li, arXiv:hep-ph/0103305, talk presented at the 4th
International Workshop on B Physics and CP violation, Ise-Shima,
Japan, Feb. 19-23, 2001; Y.Y Keum, H-n. Li, and A.I. Sanda,
arXiv:hep-ph/0201103, talk presented at the 9th international
Symposium on Heavy Flavor Physics, Caltech, Pasadena, Sept.
10-13,2001.
\bibitem{CKL} C.H. Chen, Y.Y. Keum, and H-n. Li, Phys. Rev. D {\bf 64},
112002 (2001).
\bibitem{Mi} S. Mishima, Phys. Lett. B {\bf 521}, 252 (2001).
\bibitem{CY1} H.Y. Cheng and K.C. Yang, Phys. Rev. D {\bf 64}, 074004
(2001).
\bibitem{HMW} X.G. He, J.P. Ma, and C.Y. Wu,
Phys. Rev. D {\bf 63}, 094004 (2001).
\bibitem{TLS} T. Kurimoto, H-n. Li, and A.I. Sanda, Phys. Rev. D {\bf 65},
014007 (2002).
\bibitem{WY} Z.T. Wei and M.Z. Yang, arXiv:hep-ph/0202018.
\bibitem{L4} H-n. Li, Phys. Rev. D {\bf 64}, 014019 (2001).
\bibitem{NL} M. Nagashima and H-n. Li, hep-ph/0202127.
\bibitem{CL} C.H. Chang and H-n. Li, Phys. Rev. D {\bf 55}, 5577 (1997).
\bibitem{YL} T.W. Yeh and H-n. Li, Phys. Rev. D {\bf 56}, 1615
(1997).
\bibitem{GN} A.G. Grozin and M. Neubert, Phys. Rev. D {\bf 55},
272 (1997).
\bibitem{BTF} M. Beneke and T. Feldmann, Nucl. Phys. {\bf B592}, 3
(2000).
\bibitem{DC} S. Descotes and C.T. Sachrajda, Nucl. Phys. {\bf B625},
239 (2002).
\bibitem{KK} H. Kawamura, J. Kodaira, C.F. Qiao, and K. Tanaka,
Phys. Lett. B {\bf 523}, 111 (2001); arXiv:hep-ph/0112174.
\bibitem{PB1} P. Ball, V.M. Braun, Y. Koike, and K. Tanaka, Nucl. Phys.
{\bf B529}, 323 (1998).
\bibitem{PDG} Particle Data Group, D.E. Groom {\it et al.}, Eur. Phys. J. C
{\bf 15}, 1 (2000).
\bibitem{CG} C.H. Chen and C.Q. Geng, Phys. Rev. D {\bf 63}, 114025 (2001); arXiv:hep-ph/0203003.
\bibitem{ABH} A. Ali, P. Ball, L.T. Handoko, and G. Hiller,
Phys. Rev. D {\bf 61}, 074024 (2001); A. Ali and A. Ya.
Parkhomenko, arXiv:hep-ph/0105302.
\bibitem{CY2} H.Y. Cheng and K.C. Yang, Phys. Lett. B {\bf 511}, 40
(2001).
\bibitem{CLEO} CLEO Collaboration, R.A. Briere {\it et al.},
Phys. Rev. Lett {\bf 86}, 3718 (2001).
\bibitem{Belle}  BELLE Collaboration, A. Bozek, Proceedings for
the 4th International Workshop on B Physics and CP Violation, p. 81,
Ise-Shima, Japan, Feb. 19-23, 2001.
\bibitem{Babar} BABAR Collaboration, B. Aubert {\it et al.},
Phys. Rev. Lett {\bf 87}, 151801 (2001).
\bibitem{LEP} Review of Particle Physics, Eur. Phys. J. C {\bf 3}, 1 (1998).
\bibitem{BS} J. Botts and G. Sterman, Nucl. Phys. {\bf B325}, 62 (1989).
\bibitem{LS} H-n. Li and G. Sterman Nucl. Phys. {\bf B381}, 129 (1992).
\bibitem{L5} H-n. Li, arXiv:hep-ph/0102013.
\bibitem{WYL} C.Y. Wu, T.W. Yeh, and H-n. Li, Phys. Rev. D {\bf 55}, 237
(1997).


\end{references}
\end{document}